\documentclass[conference]{IEEEtran}
\IEEEoverridecommandlockouts
\usepackage{cite}
\usepackage{amsmath,amssymb,amsfonts}
\usepackage{algorithmic}
\usepackage{graphicx}
\usepackage{textcomp}
\usepackage{xcolor}
\usepackage{url}
\usepackage{multirow}
\usepackage{caption}
\def\BibTeX{{\rm B\kern-.05em{\sc i\kern-.025em b}\kern-.08em
    T\kern-.1667em\lower.7ex\hbox{E}\kern-.125emX}}
\begin{document}

\title{Exploring an Inter-Pausal Unit (IPU) based Approach for Indic End-to-End TTS Systems}

\author{
\IEEEauthorblockN{Anusha Prakash}
\IEEEauthorblockA{\textit{Independent Researcher (Work carried out at IITM)} \\
Bengaluru, India}
\and
\IEEEauthorblockN{Hema A Murthy}
\IEEEauthorblockA{\textit{Dept of Computer Science \& Engineering} \\
\textit{Indian Institute of Technology Madras}\\
Chennai, India}
}

\maketitle

\begin{abstract}

Sentences in Indian languages are generally longer than those in English. Indian languages are also considered to be phrase-based, wherein semantically complete phrases are concatenated to make up sentences. Long utterances lead to poor training of models and result in poor prosody during synthesis. In this work, we explore an inter-pausal unit (IPU) based approach in the end-to-end (E2E) framework, focusing on synthesising conversational-style text. We consider both autoregressive Tacotron2 and non-autoregressive FastSpeech2 architectures in our study and perform experiments with three Indian languages, namely, Hindi, Tamil and Telugu. With the IPU-based Tacotron2 approach, we see a reduction in insertion and deletion errors in the synthesised audio, providing an alternative approach to the FastSpeech(2) network in terms of error reduction. The IPU-based approach requires less computational resources and produces prosodically richer synthesis compared to conventional sentence-based systems. 
\end{abstract}


\section{Introduction}
\label{sec:intro}

The initial end-to-end (E2E) architectures for text-to-speech (TTS) synthesis systems, such as Tacotron2 \cite{tacotron2}, were autoregressive based. Autoregressive networks are prone to insertion and deletion of sounds in the synthesised audio mainly due to the cross-attention between character/phoneme embeddings and mel-spectrogram frames not being learnt correctly. Unseen contexts may also result in error propagation at the output. To mainly alleviate these issues and speed up the synthesis time, non-autoregressive FastSpeech \cite{fastspeech} and FastSpeech2 \cite{fastspeech2} architectures were developed. In parallel to the development of these non-autoregressive architectures, we have explored an inter-pausal unit (IPU) based approach, previously proposed in the hidden Markov model (HMM) based framework \cite{Jeena_PhraseJournal_2019}, to primarily address the issue of error-prone synthesis in autoregressive E2E networks. Specifically, the focus is on synthesising conversational-type text, such as classroom lectures. 

Indian language sentences are much longer compared to those in English. Moreover, Dravidian languages are characterised by agglutination \cite{AnushaIS16}, where multiple words are concatenated together to form a single word. It has been observed that if the training data has, on average, long utterances (greater than 20 seconds in duration), the Tacotron2 model is not learnt well. The alignment between text and speech frames may not be learnt correctly, despite including a monotonicity constraint \footnote{The alignment between phone and acoustic sequences is monotonic.}. Long utterances also require more computational memory to accommodate large batch sizes and may result in out-of-memory issues during training. If the attention module does not learn the alignments correctly, then the end-of-utterance may not be predicted correctly during synthesis. The end-of-utterance may be predicted at an earlier decoder step, leading to skipping of words in the synthesised audio. Or the end-of-utterance may be predicted after the actual number of mel-spectrogram frames, leading to the repetition of words. This is especially evident in the synthesis of long sentences and texts that are conversational in nature. Consider the transcribed text corresponding to a classroom lecture. The text corresponds to spontaneous speech, which is unrehearsed and may contain grammatically incomplete utterances. Classical machine learning approaches handle unseen contexts using tree-based clustering or suitable smoothening techniques. Coupled with the attention not being learnt correctly, auto-regressive networks may be error-prone while synthesising unseen contexts. A TTS system, trained on complete sentences, may fail to correctly synthesise the conversational-style text. Even if a long text is synthesised correctly, the synthesised audio may suffer from poor prosody. Hence, it is necessary to split the long sentences into shorter segments, both during training and synthesis.

\cite{Jeena_PhraseJournal_2019} explored an inter-pausal unit (IPU) based approach in the HMM framework, which led to better synthesis compared to the performance of conventional sentence-based systems. Motivated by this, we propose to explore the IPU-based approach in the context of E2E systems. The training data, which is available at the sentence level, is first aligned and spliced at intra-utterance silence regions based on a certain threshold. These spliced segments are referred to as inter-pausal units (IPUs). A TTS voice is trained on the obtained IPU data in an E2E framework and compared with a system trained using sentence-level data. During synthesis, IPU-based segments are individually passed to the model and synthesised instead of the entire sentence being passed to the model. Finally, the individual segments are concatenated to generate the complete synthesised audio. We analyse the following scenarios in this study:

\begin{itemize}
    \item We see how the IPU-based approach helps reduce errors in the Tacotron2 framework. A significant advantage is the reduction in the training time of the model by more than $50\%$ with the IPU-based approach compared to the sentence-based method.
    \item By initialising a sentence-based system with a pre-trained IPU-based system, we get faster convergence at early epochs.
    \item We see that the IPU-based Tacotron2 model is comparable to a sentence-based non-autoregressive FastSpeech2 system in terms of error reduction. However, the IPU-based approach still helps in generating prosodically richer audio by explicitly resetting the prosodic parameters at the beginning of each IPU segment.
\end{itemize}

The organisation of this paper is as follows. An analysis of utterance duration across various datasets is presented in Section \ref{sec:chap5-utts_len}. Section \ref{sec:chap5-splitting} discusses the importance of splitting sentences into shorter segments and the significance of using IPUs. Section \ref{sec:chap5-related} presents related work in the literature, with a brief review of the IPU-based work in the HMM framework. The baseline sentence-based and IPU-based approaches are described in Sections \ref{sec:chap5-baseline} and \ref{sec:chap5-proposed_IPU}, respectively. The experimental set-up and the various analyses are described in Section \ref{sec:chap5-experiments}. Section \ref{sec:summary} summarises the work.

\section{Analysis of utterance length across various datasets}
\label{sec:chap5-utts_len}

Most studies that propose new architectures or advancements are conducted mainly with English datasets. Papers on Tacotron2, FastSpeech(2), WaveNet and WaveGlow use the LJSpeech dataset \cite{ljspeech17} in their experiments. The utterances in this dataset are between 1-10 seconds in duration. Models trained with other datasets and other languages may have their own unique challenges, and standard procedures to train TTS systems with existing architectures may need re-visiting.

\begin{table}[h!]
\centering
\caption{Datasets considered for utterance length analysis}
\label{tab:chap5-dataset_label}
\begin{tabular}{|c|l|c|}
\hline
\textbf{Dataset label} & \multicolumn{1}{c|}{\textbf{Dataset}} & \textbf{Language} \\ \hline
A                      & Aishell3 \cite{AISHELL_3_2020}                              & Mandarin          \\ \hline
B                      & LJSpeech \cite{ljspeech17}                              & English           \\ \hline
C                      & VCTK \cite{veaux2017cstr_vctk}                                  & English           \\ \hline
D                      & Siwis French dataset \cite{SIWIS}                                 & French            \\ \hline
E                      & Indic Hindi male \cite{ArunResources2016}                      & Hindi             \\ \hline
F                      & Indic Tamil female \cite{ArunResources2016}                    & Tamil             \\ \hline
G                      & Indic Telugu male \cite{ArunResources2016}                     & Telugu            \\ \hline
\end{tabular}
\end{table}

\begin{figure}[!h]
 \centering
 \includegraphics[width = \linewidth]{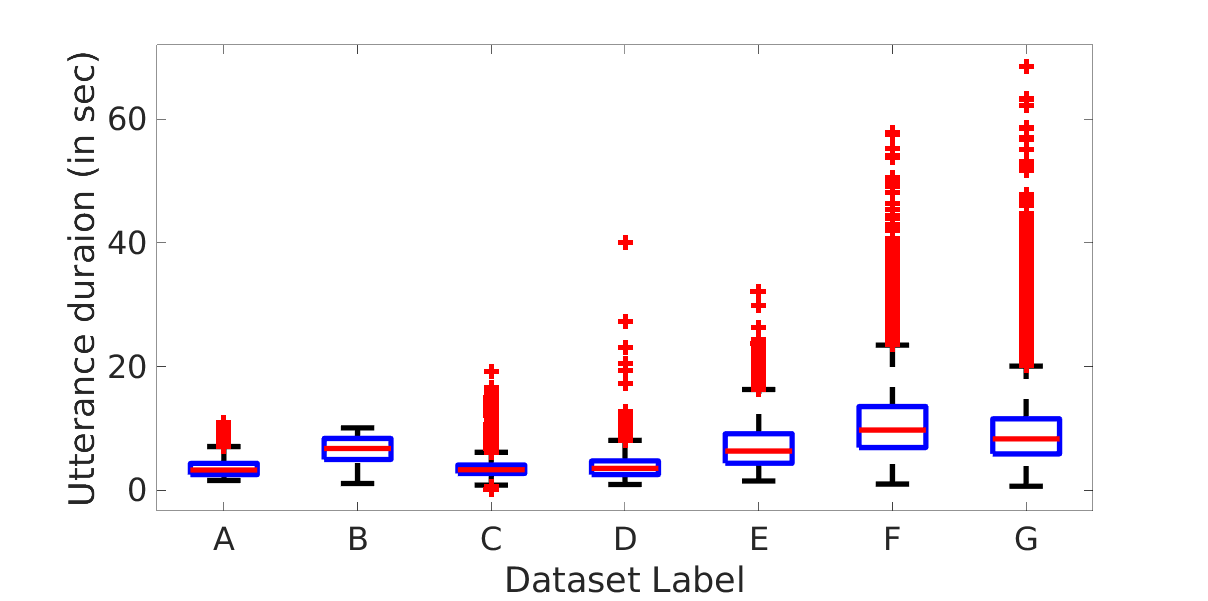}
 \caption{Distribution of length of utterances across various datasets. The dataset label identifier is given in Table \ref{tab:chap5-dataset_label}.}
  \label{fig:chap5-utts_dur_box}
\end{figure}

We analyse the utterance duration across various commonly used datasets for TTS training. The datasets under consideration are listed in Table \ref{tab:chap5-dataset_label}. Datasets E, F and G are Indian language datasets \cite{ArunResources2016}. The distribution of the duration of utterances across the datasets is shown as box plots in Figure \ref{fig:chap5-utts_dur_box}. It is clearly seen that utterances of Indian languages are longer compared to those of other languages/datasets. Utterances of Tamil and Telugu are especially longer due to the agglutinative nature of their scripts.

\section{Splitting utterances into shorter segments}
\label{sec:chap5-splitting}

The alignment between character sequences and acoustic features is learnt more effectively on shorter sequences \cite{attention_NIPS2015}. Hence, it is better to split the data into shorter segments for training. Randomly splitting the data into shorter segments may destroy the prosodic patterns at sentence and discourse levels. Consider the sentence, ``The panda eats shoots and leaves''.  The position of commas in this sentence determines the meaning of the sentence, as given below:
\begin{itemize}
    \item \textit{The panda eats shoots and leaves}-- this means that the panda is eating shoots and leaves
    \item \textit{The panda eats, shoots and leaves}-- this refers to three actions of the panda, namely, eating, shooting and leaving.
\end{itemize}

It is vital to split the utterances into more natural segments. Indian language utterances are made of a sequence of phrases rather than a sequence of sentences. Studies by \cite{fery2010intonation} on Bengali, Hindi, Malayalam, and Tamil classify Indian languages into intonational phrase languages. Phrases are also generally semantically complete. Hence, as a design choice, phrases could be a better alternative to sentences for Indian languages.

The easiest approach to text phrasing would be to consider punctuation marks in the text as phrase boundaries. However, texts in Indian languages are rarely punctuated. This is because Indian languages are, to a large extent, word-order free, i.e., the order of words contains only secondary information \cite{Bharati_wordOrder}. Traditionally, Indian language texts did not have any punctuation marks, including commas, spaces and full stops \cite{hellwig-2016-detecting}. Parts-of-speech (POS) taggers for Indian languages are still at a nascent stage and are available for very few Indian languages. Studies performed by \cite{raghavThesis} show that a sentence can be phrased in multiple ways, and it could vary with speakers. Hence, utterances
corresponding to the training data are spliced wherever the speaker has paused (IPUs) \cite{Jeena_PhraseJournal_2019}. The TTS systems are then trained with IPUs and compared with conventional sentence-based systems. Studies by \cite{pitch_reset} show that the pitch is reset at the beginning of a phrase. By training on IPUs, we attempt to ensure this indirectly.

\section{Related work}
\label{sec:chap5-related}

Most existing works remove long utterances beyond a threshold before training the E2E models. In this section, we present related literature in the context of handling long utterances in the training data.

In the Blizzard Challenge 2019 \cite{blizzard2019}, participants were tasked with training a TTS system given an 8-hour Mandarin dataset. The dataset consisted of 60-second audio clips from a Chinese talk show program. For this task, \cite{blizzard2019_mobvoi, blizzard2019_dku, blizzard2019_stc, blizzard2019_tju, blizzard2019_imu, blizzard2019_sjtu, blizzard2019_sznpu, blizzard2019_cmu, blizzard2019_vivo} proposed different approaches to handling long training utterances. In \cite{blizzard2019_mobvoi}, long audio segments were spliced into shorter segments of a maximum of 10 seconds in duration. Other works \cite{blizzard2019_dku, blizzard2019_stc, blizzard2019_tju} obtained shorter segments based on punctuations in the text and using forced alignment. \cite{blizzard2019_imu} and \cite{blizzard2019_sjtu} employed voice activity detection (VAD) tools for splicing, with mistakes in VAD being manually corrected. \cite{blizzard2019_cmu} proposed a sub-sentence training by splicing the utterances into shorter segments between 5-23 seconds. The criterion for splicing is not clearly described in the paper. Further, \cite{blizzard2019_cmu} reported that one of the main contributing factors to reduced system performance was sub-sentence training. In \cite{blizzard2019_vivo}, utterances were segmented into shorter audio files by considering various thresholds, including pause duration. However, a few utterances terminated in the middle of syllables, which had to be manually checked and removed from the training data.

There are also two parallel works \cite{cong2020ppspeech, lenglet21_ssw} which deal with long training utterances in the Tacotron2 architecture. The focus of \cite{lenglet21_ssw} is to study the impact of segmentation on speech quality and expressiveness. \cite{lenglet21_ssw} use the  M-AILABS French dataset \cite{mailabs}, where the utterance duration is between 1-20 seconds. Utterances are spliced at pauses $>$ 400 msec, with close to $95\%$ of them coinciding with punctuation marks in the text. Then a Tacotron2 model is trained on the shorter audio segments. \cite{lenglet21_ssw} report that data segmentation impacts both speech quality and expressiveness in opposite ways and needs further investigation. In \cite{cong2020ppspeech}, intonational phrase boundaries are identified using a phrase detector trained on features, including POS taggers. Based on the boundary information, utterances are spliced into phrases for training. Context embedding is also included during training to capture prosody across neighbouring phrases. During the synthesis of a long text, phrases are identified and are individually synthesised in parallel. As a result, the synthesis time of the proposed system is less compared to that of the sentence-based Tacotron2 system.

\subsection{IPU based approach in HMM-based TTS systems}
\label{sec:chap2-IPU}

In the context of phrase-level TTS systems, studies and experiments have been carried out by \cite{Jeena_prosody_2016, Jeena_PhraseJournal_2019} in the HMM framework. These works have analysed the length of inter-pausal units in terms of syllable count for multiple Indian languages and Indian English of different nativities. They observed that the syllable count follows mostly a Gamma distribution, irrespective of speaker or language. \cite{Jeena_PhraseJournal_2019} conjecture that IPUs reflect the physiological (constraints due to respiratory requirements) and neurophysiological (speech planning) aspects of the speech production mechanism. Based on this, IPUs have been considered to be more natural segments compared to sentence-level data for training. Specifically, three TTS systems at the IPU level have been trained, corresponding to the first, middle and last IPUs. During synthesis, the test sentence is split into phrases, synthesised by the corresponding IPU model and then concatenated. The splitting into phrases is based on punctuation marks in the text, if available. In the absence of any punctuation marks, a phrase prediction technique is proposed. The prediction is based on the number of syllables in the test sentence and the Gamma distribution of the number of syllables in different IPUs, in conjunction with case markers or word terminal syllables. \cite{Jeena_PhraseJournal_2019} show that systems trained with the IPU approach outperform the sentence-based TTS systems.

Based on the literature work, we propose to use the IPU-based approach in the context of E2E systems. Preliminary experiments were performed by training categorised IPU-level E2E systems, similar to experiments in \cite{Jeena_PhraseJournal_2019}. However, the model training was poor due to the lack of adequate data. This result is in keeping with the observations in \cite{Jeena_PhraseJournal_2019}  on conventional neural-network based systems. Since the aim was to split the long sentences into smaller units, the IPUs were used to train a single IPU-based TTS voice.

The focus of our work is on systematically analysing the robustness across IPU and sentence-based systems with the Tacotron2 architecture. We also highlight the efficacy of this approach for conversational-type text, since conversational speech is made of a sequence of phrases (which could be incomplete). For purposes where a sentence-based system is necessary, we initialise its training with an IPU-based system. Most importantly, we highlight the faster convergence of such an initialised sentence-based system, thus utilising minimal computational resources. Lastly, we explore the IPU-based approach in the context of FastSpeech2, a more recent architecture.

\section{Baseline system}
\label{sec:chap5-baseline}

\begin{figure}[h!]
 \centering
 \includegraphics[width = \linewidth]{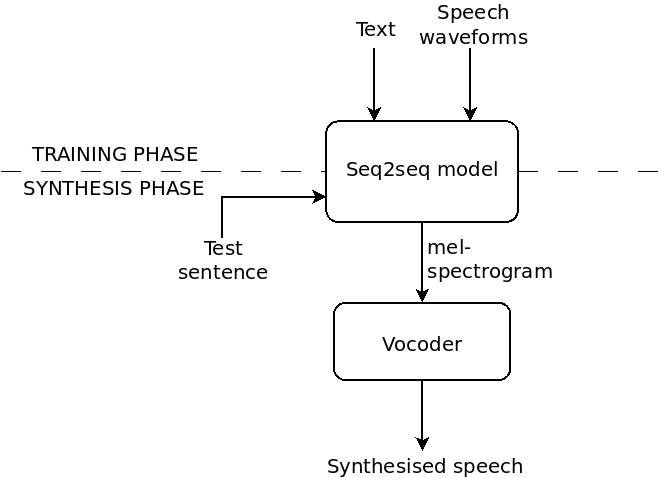}
 \caption{Training and synthesis phases of the baseline sentence-based system}
 \label{fig:chap5-baseline}
\end{figure}

The baseline system for comparison is sentence-based. The utterances in the data are not passed through any pre-processing module, and $<$text, audio$>$ pairs are directly fed to the Tacotron2/FastSpeech2 network for training. For synthesis, the test sentence is directly passed through the trained network, and the generated mel-spectrogram is fed to a vocoder. The training and synthesis phases of the baseline system are illustrated in Figure \ref{fig:chap5-baseline}.

\section{IPU based E2E training}
\label{sec:chap5-proposed_IPU}

\begin{figure}[h!]
 \centering
 \includegraphics[width = \linewidth]{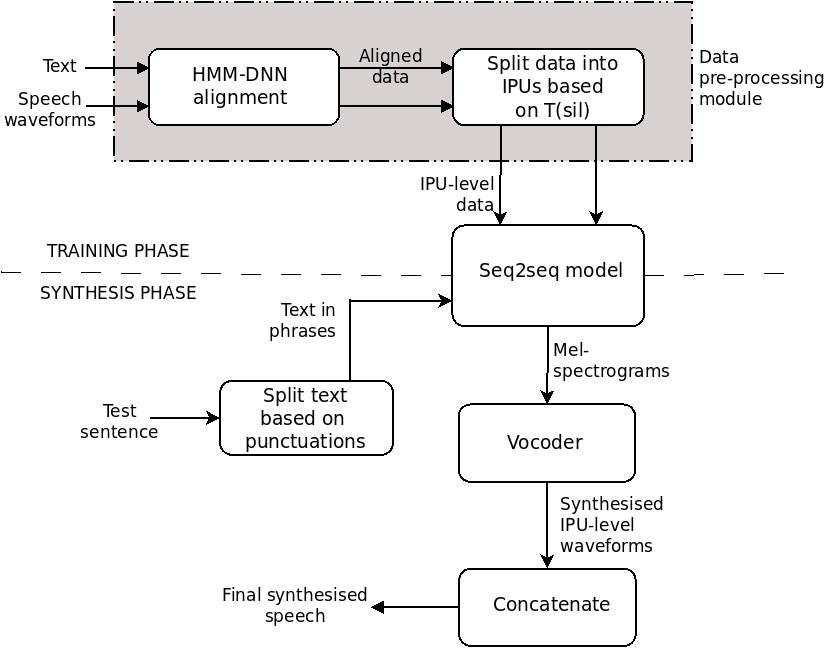}
 \caption{Training and synthesis phases of the IPU-based approach}
 \label{fig:chap5-IPU_flowchart}
\end{figure}

In contrast to the sentence-based approach, the IPU-based approach pre-processes the TTS data before training and synthesising the evaluation text. An illustration of the same is given in Figure \ref{fig:chap5-IPU_flowchart}. The training data consists of audio waveforms and their corresponding text. The IPU-level data is extracted similarly to the method proposed in \cite{Jeena_PhraseJournal_2019}. To splice the speech data into IPUs, the text or the phone sequence should be first aligned to the waveform. For this, the hybrid segmentation technique is used \cite{BABY202010}. As the training text does not have any punctuations, intra-utterance pauses need to be identified. Initially, a ``short pause'' (\textit{sp}) label is forcibly included after every word. In the first stage of alignment, group delay based cues are combined with HMM-based alignments to give the initial set of alignments. In this process, very short \textit{sp} segments ($<$ 20 msec) are identified and eliminated. The alignments are then fine-tuned using a DNN-based approach. The remaining \textit{sp} segments are considered to be the final intra-utterance silence regions.

The IPUs are extracted based on the intra-utterance silence regions. If the duration of an intra-utterance silence segment is greater than or equal to a threshold, then this segment acts as an IPU boundary. This threshold is termed as $T(sil)$. The aligned waveforms and text are split at these identified boundaries to obtain the final IPU-level data. The IPU-based  $<$text, audio$>$ pairs are directly fed to the seq2seq network for training.

During synthesis, the test sentence is first split into IPUs based on  ``commas'' or other punctuation marks. Then each IPU is passed through the seq2seq model and a neural vocoder to obtain IPU-level synthesised speech. Synthesised IPUs are concatenated in order to generate the synthesised audio at the sentence level. Punctuation marks provided in the text or by a suitable phrase predictor can be used to obtain IPUs for synthesis.

\section{Experiments and results}
\label{sec:chap5-experiments}

Datasets used in the experiments are part of the Indic TTS database \cite{ArunResources2016}. Hindi male (8.5 hours), Tamil female (10 hours) and Telugu male (10 hours) datasets are considered for the experiments. Hindi is an Indo-Aryan language, while Tamil and Telugu are Dravidian languages. Audio files are downsampled to 22.05 kHz to ensure uniformity in the feature extraction part. The corresponding text is in UTF-8 format. $90\%$ of the data is considered the training data, and the remaining $10\%$ as validation data. For training Tacotron2 and FastSpeech2 models, the ESPNet toolkit is used \cite{espnet}, with the default parameters. HiFi-GAN v1 models are trained using an open-source code \cite{hifigan_NEURIPS2020} \footnote{\url{https://github.com/jik876/hifi-gan}}.

To split the utterances into IPU segments, we consider $T(sil)$= 100 msec. This is based on the experiments in \cite{Jeena_PhraseJournal_2019}. We also present auxiliary observations on the choice of $T(sil)$ towards the end of this section.

\subsection{Sentence vs. IPU-based TTS (Tacotron2) systems}

Sentence-based and IPU-based Tacotron2 models are trained for Hindi, Tamil and Telugu datasets as described in Sections \ref{sec:chap5-baseline} and \ref{sec:chap5-proposed_IPU}. The sentence-based Tamil and Telugu systems train poorly, mainly due to the attention not being learnt correctly on long training utterances (refer to Figure \ref{fig:chap5-utts_dur_box}). In comparison, their IPU-based counterparts get trained well. The Hindi sentence-based TTS system is trained well and is used to compare with the corresponding IPU-based system. Hence, most of the analysis in this sub-section is focused on the Hindi dataset. In the subsequent experiments, we will revisit the other datasets for analysis and evaluation.

\subsubsection{Utterance length analysis of the training data}

\begin{figure}[!h]
 \centering
 \includegraphics[width = \linewidth]{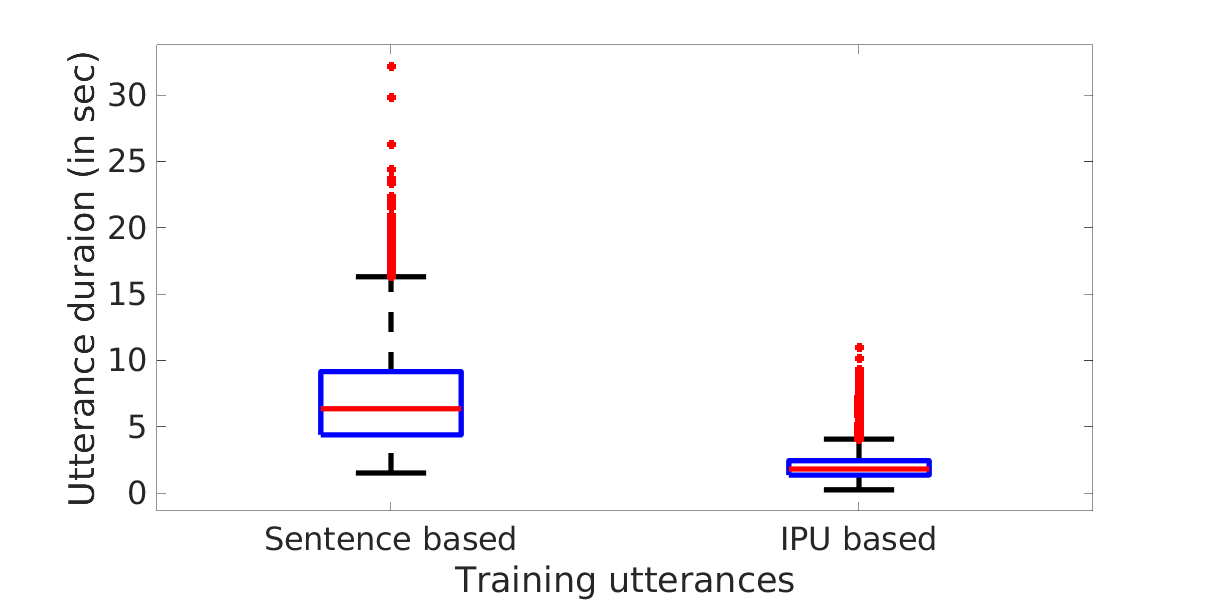}
 \caption{Distribution of length of training utterances across sentence and IPU-based systems (Hindi male dataset)}
  \label{fig:chap5-IPUutts_dur_hin}
\end{figure}

\begin{figure}[!h]
 \centering
 \includegraphics[width = \linewidth]{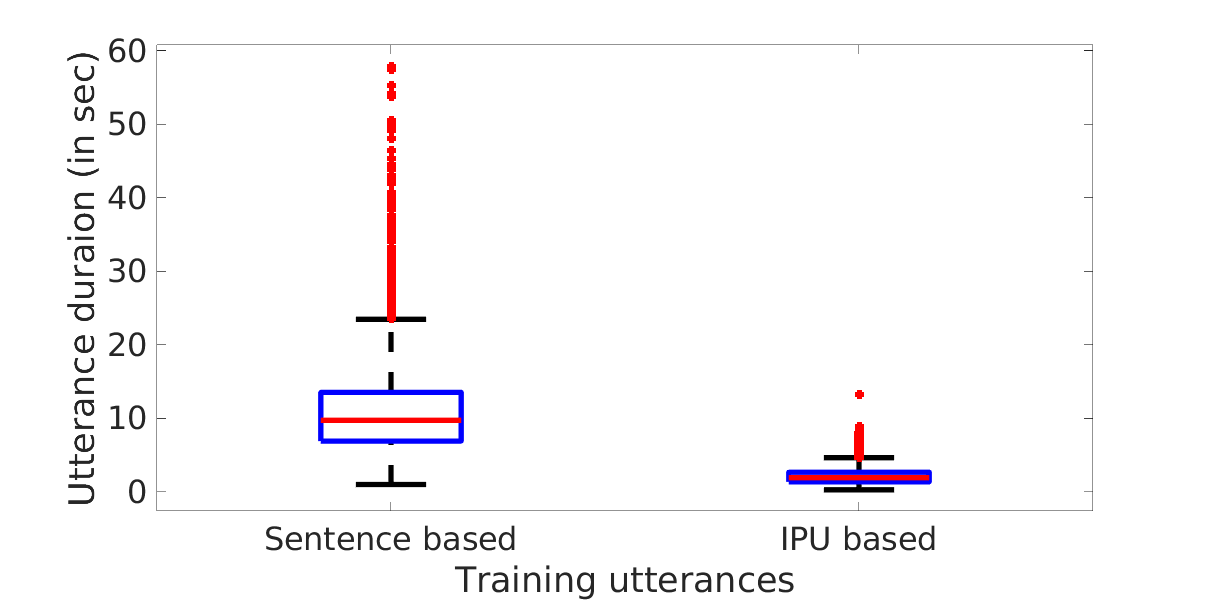}
 \caption{Distribution of length of training utterances across sentence and IPU-based systems (Tamil female dataset)}
  \label{fig:chap5-IPUutts_dur_tam}
\end{figure}

 Figures \ref{fig:chap5-IPUutts_dur_hin} and \ref{fig:chap5-IPUutts_dur_tam} show the utterance length comparison of the training data across sentence and IPU-based approaches for Hindi and Tamil, respectively. With the IPU-based approach, the mean and range of the training utterance durations are reduced. This leads to the training of better models with the alignments/attention restricted within the IPU segment.

\subsubsection{Error analysis}

To analyse the robustness of the trained models, a set of conversational-style text is synthesised by different TTS systems. The generated audio files are checked for errors, such as word skips and repetitions. This analysis is performed manually. Text corresponding to classroom lectures from the National Programme on Technology Enhanced Learning (NPTEL) \cite{nptel} is considered.

An entire lecture is split into smaller segments based on VAD and then transcribed using an automatic speech recogniser (ASR). It is not guaranteed that the text always terminates with a meaningful or complete sentence/phrase. Then, using an English-to-Hindi machine translation system, the text is translated to Hindi for synthesis. At each stage, manual verification and correction are conducted, as described in \cite{v2v_IS23}.

Four NPTEL lectures on subjects belonging to different disciplines are considered-- Basic Electronics (BE, Electrical Engineering), Cloud Computing (CC, Computer Science), Language and Mind (LM, Humanities), and Product Design and Development (PDD, Mechanical Engineering). 100 translated text segments from each lecture are then presented for synthesis. This analysis is only carried out for Hindi due to the availability of corrected Hindi translations across various subjects \footnote{Synthesised samples for Hindi are available at \url{https://nltm.iitm.ac.in/IPU_based_TTS/index.html}}.

The conversational-style text is passed directly to the sentence and IPU-based systems. It is to be noted that the test text does not contain any punctuation marks and is hence not split into smaller segments for the IPU-based TTS, as described in the synthesis phase of Figure \ref{fig:chap5-IPU_flowchart}.
The error analysis is presented in Table \ref{tab:chap5-error_analysis_Tac2}. It is clearly seen that the sentence-based TTS system generates audio that has repetitions in many cases and word skips in a few instances. The repetition and skip errors are $28.25\%$ and $2.5\%$, respectively. In contrast, the repetition errors with the IPU-based system have reduced to $0\%$. However, the IPU-based system still has a word skip error of $1.5\%$. Word skips are mainly observed for words that generally follow the end of a phrase.

% In contrast, the audio files generated by the IPU based TTS does not have any repetitions; however word skips are observed in a few instances. 

% \usepackage{multirow}
\begin{table}[]
\centering
\caption{Analysis of category-wise errors across sentence and IPU-based Tacotron2 systems on conversational-style Hindi text from different subjects}
\label{tab:chap5-error_analysis_Tac2}
\begin{tabular}{|c|l|r|r|r|r|r|}
\hline
\textbf{TTS System}             & \multicolumn{1}{c|}{\textbf{\begin{tabular}[c]{@{}c@{}}Error\\ type\end{tabular}}} & \multicolumn{1}{c|}{\textbf{\begin{tabular}[c]{@{}c@{}}BE\\ \end{tabular}}} & \multicolumn{1}{c|}{\textbf{\begin{tabular}[c]{@{}c@{}}CC\\\end{tabular}}} & \multicolumn{1}{c|}{\textbf{\begin{tabular}[c]{@{}c@{}}LM\\ \end{tabular}}} & \multicolumn{1}{c|}{\textbf{\begin{tabular}[c]{@{}c@{}}PDD\\ \end{tabular}}} & \multicolumn{1}{c|}{\textbf{Total}} \\ \hline
\multirow{2}{*}{Sentence} & Repetitions                                 & 30 & 26                                            & 25 & 32 & 113 \\
\cline{2-7} & Skips & 1 & 4 & 2 & 3 & 10        \\ \hline
\multirow{2}{*}{IPU}      & Repetitions                                 & 0 & 0  & 0 & 0  & 0                                   \\ \cline{2-7} 
    & Skips  & 2  & 0  & 1  & 3  & 6                                   \\ \hline
\end{tabular}
\end{table}

\begin{figure}[!h]
 \centering
  \captionof{table}{Sample Hindi text with English translations erroneously generated by the sentence-based Tacotron2 Hindi system. The red and blue texts indicate words repeated and skipped, respectively.}
  \label{tab:chap5-sent_err_text}
 \includegraphics[width = \linewidth]{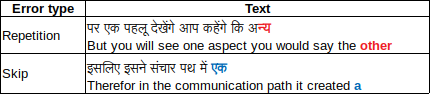}
\end{figure}

Table \ref{tab:chap5-sent_err_text} shows sample Hindi texts that were erroneously synthesised by the Hindi sentence-based Tacotron2 model. Both texts are grammatically incomplete.

\begin{figure}[!h]
 \centering
  \captionof{table}{Sample phrase breaks identified from Hindi and Tamil IPU data}
  \label{tab:chap5-phrase_breaks}
 \includegraphics[width = 0.6\linewidth]{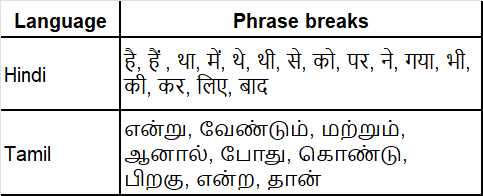}
\end{figure}

To address the issue of word skips in the IPU-based TTS system, the synthesis phase is as described in Figure \ref{fig:chap5-IPU_flowchart}. The text is split into smaller segments or phrases before synthesis. This is a challenge due to the absence of any punctuation marks in the text. Hence, IPU breaks are identified from the training data. The words appearing at the end of each IPU segment in the text are noted and sorted based on the frequency of occurrence. Then the most frequent words in the list are considered phrase/IPU breaks. Sample phrase breaks for Hindi and Tamil are given in Table \ref{tab:chap5-phrase_breaks}. We also include an additional constraint on the length of the IPU segment. If an IPU has less than 3 words, then the unit is merged with the next IPU. This is an empirical design choice and can be fine-tuned for other datasets.

Alternatively, phrase breaks in the text can also be identified by building a classification and regression tree (CART) as done in \cite{klimkov17_interspeech} or using the phrase prediction technique proposed in \cite{Jeena_PhraseJournal_2019}. A POS tagger can also be used if available. This kind of splitting into smaller segments for synthesis also helps reduce word skip errors for sentence-based TTS systems, but does not considerably reduce the repetition errors.

\subsubsection{End-of-utterance prediction}
\vspace{1cm}

\begin{figure}[!h]
 \centering
 \includegraphics[width = \linewidth]{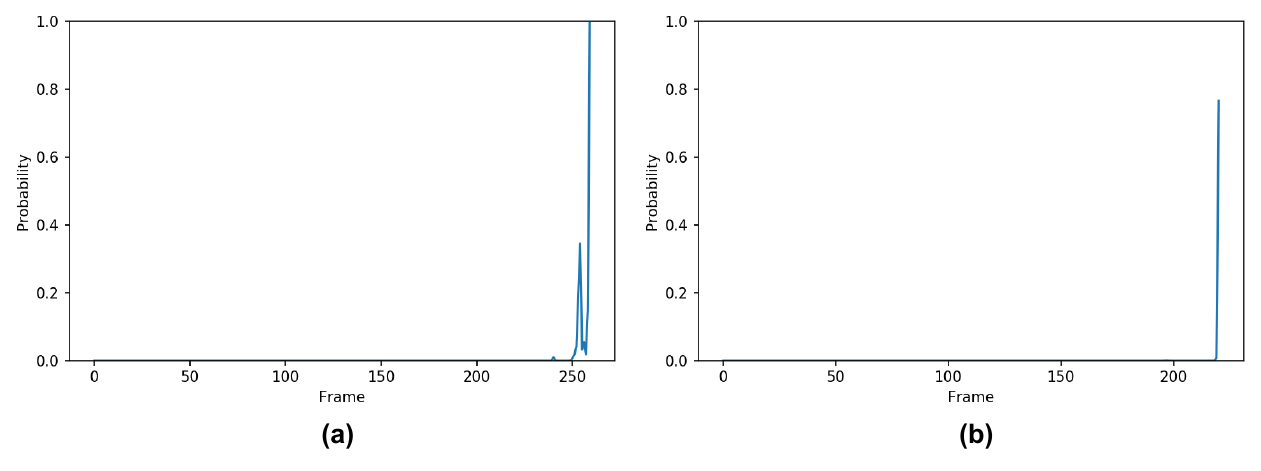}
 \caption{End-of-utterance prediction corresponding to a sample Hindi text synthesised by (a) sentence-based (b) IPU-based TTS systems}
  \label{tab:chap5-EOS_sent_IPU}
\end{figure}

We also analyse how the end-of-utterance ($<$eou$>$) prediction affects the synthesis output. Consider the plots given in Figure \ref{tab:chap5-EOS_sent_IPU}. The x-axis corresponds to the mel-spectrogram frames, and the y-axis indicates the probability that the system has completed generation. In this example, the IPU-based TTS model generates audio correctly with respect to the presented Hindi text. The $<$eou$>$ is predicted around the 230th frame (Figure \ref{tab:chap5-EOS_sent_IPU} (b)). However, the sentence-based TTS predicts the $<$eou$>$ after the 250th frame (Figure \ref{tab:chap5-EOS_sent_IPU} (a)) for the same text. These extra mel-spectrogram frames generated by the sentence-based TTS voice appear as repetitions in the synthesised audio. Based on the error analysis presented previously, the sentence-based system makes incorrect $<$eou$>$ predictions $30.75\%$ of the time. We also train a language model (with a maximum order of 3) on the training data. We then determine the log-likelihood scores of texts presented to the sentence-based Hindi model for synthesis. The average sentence-level log-likelihood scores corresponding to 10 correctly and 10 incorrectly synthesised utterances are $-34.81$ and $-43.96$, respectively. The lower likelihood score of the text could be a contributing factor for the auto-regressive network to predict $<$eou$>$ incorrectly.

\subsubsection{Subjective evaluation}

We conduct a subjective pairwise comparison (PC) test comparing the performances of sentence and IPU-based Hindi systems. Tamil and Telugu systems are excluded from this test as their sentence-based systems clearly do not generate the required output due to poor training. 

In the PC test, listeners are presented (in random order) with the audio synthesised by both systems for the same text. Evaluators are required to rate their preference for each system. From the set considered for error analysis, 20 synthesised utterances are randomly chosen. For each chosen utterance, the previous or successive utterance(s) are concatenated such that the meaning to be conveyed is complete. This is to avoid any bias by the evaluators with respect to incomplete content. 10 native Hindi speakers participated in the evaluation, with each listener evaluating 10 audio pairs. Results of the PC test are presented in Table \ref{tab:chap5-pc_exp1}. It is clearly seen that the IPU-based system is preferred over the conventional sentence-based system. The difference in performance of both systems is statistically significant ($p < 0.05$).

\begin{table}[h!]
\centering
\caption{PC test results comparing Hindi sentence and IPU-based (Tacotron2) systems}
\label{tab:chap5-pc_exp1}
\begin{tabular}{|r|r|r|r|}
\hline
\multicolumn{1}{|c|}{\textbf{System}} & \multicolumn{1}{|c|}{\textbf{Sentence}} & \multicolumn{1}{c|}{\textbf{IPU}} & \multicolumn{1}{c|}{\textbf{Equal}} \\ \hline
\textbf{Preference (in $\%$}) &   12.00   &     80.50        &      7.50                          \\ \hline
\end{tabular}
\end{table}

\subsubsection{Comparison of training time}

Table \ref{tab:chap5-training_time} shows the average training time per epoch for the sentence and IPU-based Tacotron2 systems. An average relative reduction in training time by  $59.60\%$ is obtained with the IPU approach. This demonstrates that the IPU-based approach requires less GPU resources and is hence computationally less intensive.

\begin{table}[h!]
\centering
\caption{Comparison of training time across different Tacotron2 systems}
\label{tab:chap5-training_time}
\begin{tabular}{|l|cr|}
\hline
\multicolumn{1}{|c|}{\multirow{2}{*}{Dataset}} & \multicolumn{2}{c|}{Training time per epoch (min)}                   \\ \cline{2-3} 
\multicolumn{1}{|c|}{}                         & \multicolumn{1}{c|}{Sentence-based} & \multicolumn{1}{c|}{IPU-based} \\ \hline
Hindi male                                     & \multicolumn{1}{r|}{5.36}           & 2.18                           \\ \hline
Tamil female                                   & \multicolumn{1}{r|}{6.23}           & 2.50                            \\ \hline
\end{tabular}
\end{table}

In a nutshell, we observe the following advantages of the IPU-based system over the conventional sentence-based Tacotron2 system:

\begin{itemize}
    \item We are able to reduce the length of the training utterances with the IPU approach, leading to better training of the models.
    \item With better learning of the alignments across text embedding and mel-spectrogram frames, the $<$eou$>$ modelling is improved.
    \item The number of repetition and word skip errors are reduced (to almost zero) with the IPU-based approach, especially by splitting the test text at the IPU breaks.
    \item The IPU system has a faster training time and is hence computationally less expensive.
\end{itemize}

\subsection{Initialisation of sentence based (Tacotron2) TTS with IPU based TTS}

\begin{figure}[!h]
 \centering
 \includegraphics[width = \linewidth]{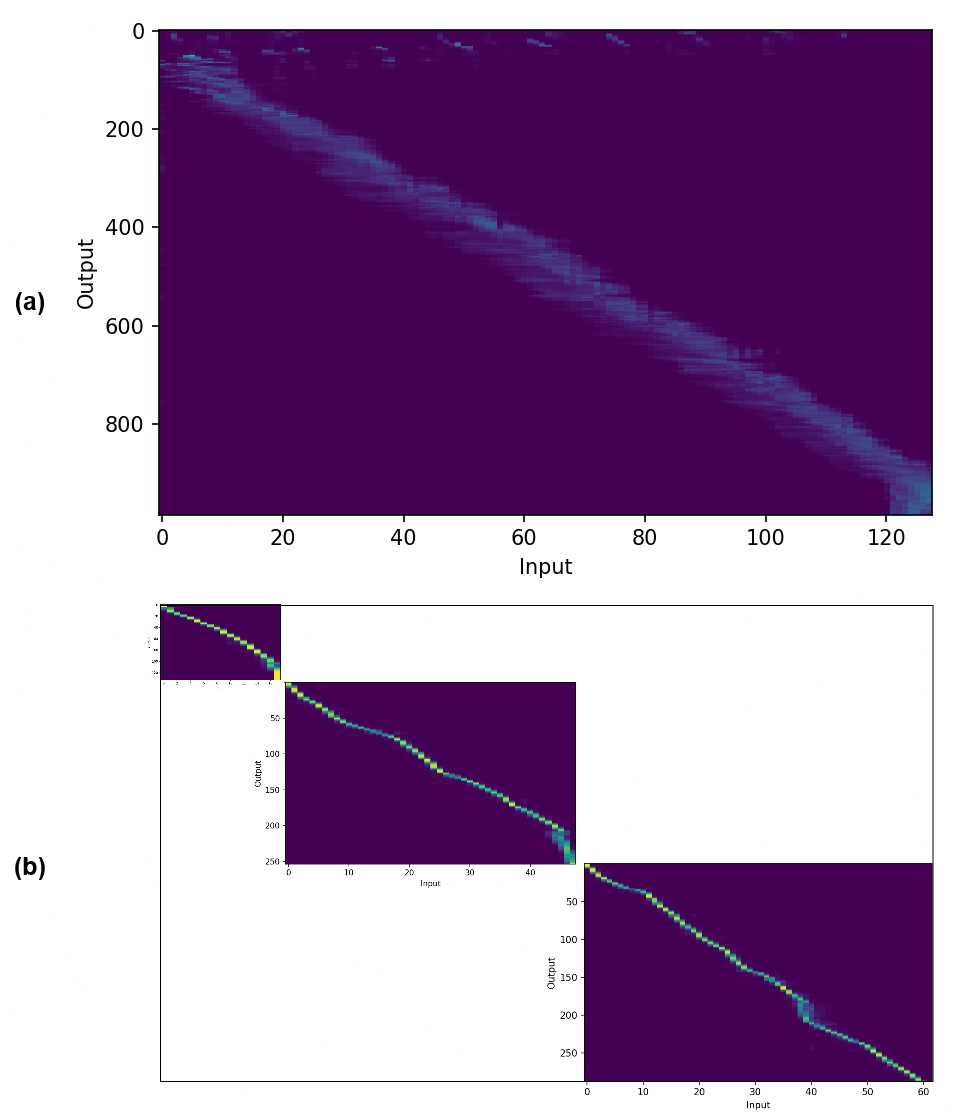}
 \caption{Attention plots corresponding to a sample Telugu text with the: (a) conventional sentence-based Telugu TTS system, (b) IPU-based Telugu TTS system on individual IPU segments.}
  \label{fig:chap5-tel_attnPlots_exp1}
\end{figure} 

In the sentence-based approach, the alignments are modelled using the attention module, considering the entire utterance. If the training utterances are very long, then attention is learnt poorly, as seen in the case of Tamil and Telugu systems. With the IPU-based system, the attention is restricted to the IPU level having shorter segments, and is learnt properly. The attention plots corresponding to a sample Telugu text are shown in Figure \ref{fig:chap5-tel_attnPlots_exp1}. The example text has 3 IPU segments. While the Telugu sentence-based system struggles to learn the attention across the entire sentence/utterance (as evidenced by the blurry attention of Figure \ref{fig:chap5-tel_attnPlots_exp1}(a)), the IPU-based system learns almost correct alignments within the constituent IPU segments (Figure \ref{fig:chap5-tel_attnPlots_exp1}(b)).

Now the question is, can we initialise a sentence-based TTS system with an IPU-based system? And if we can successfully achieve this objective, how is the performance of the initialised sentence-based system compared to the conventional sentence-based system? We investigate these scenarios with the Telugu data. We retrain a sentence-based Telugu TTS system from its corresponding IPU-based system. We analyse system performance in terms of the convergence time and subjective PC test.

\subsubsection{Convergence} 

\begin{figure}[!h]
 \centering
 \includegraphics[width = \linewidth]{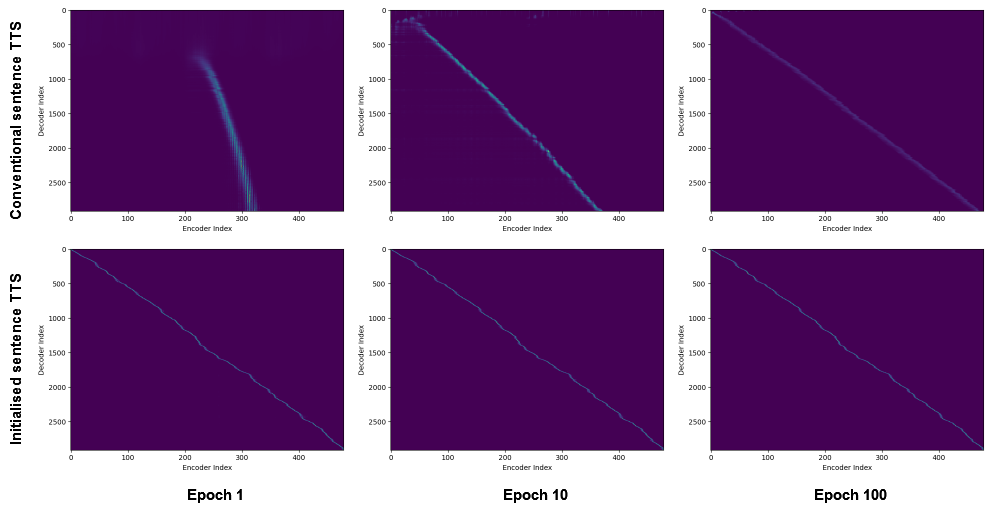}
 \caption{Attention plots corresponding to conventional and IPU-initialised sentence-based TTS systems for Telugu}
  \label{fig:chap5-tel_attnPlots}
\end{figure}

Figure \ref{fig:chap5-tel_attnPlots} shows the attention plots of different training epochs across the two sentence-based Telugu systems. It is clearly seen that the IPU initialisation helps in achieving convergence (indicated by the nearly diagonal attention) at early epochs itself. In contrast, the conventional sentence-based system struggles to reach convergence; even at the 100th epoch, the attention plot is blurred.

\subsubsection{Subjective evaluation}

We conduct a PC test comparing the performances of IPU-initialised sentence-based and IPU-based Telugu Tacotron2 systems. Translated text segments of a classroom lecture on the subject ``Artificial Intelligence'' from SWAYAM \cite{swayam} are synthesised. From this set, 20 utterances are randomly selected, and the previous and/or successive utterance(s) are concatenated such that the meaning to be conveyed is complete. 

% \begin{landscape}
% \end{landscape}

11 native Telugu speakers participated in the evaluation, with each listener evaluating 10 audio pairs. Results of the PC test are presented in Table \ref{tab:chap5-pc_exp2_tel}. It is clearly seen that the IPU-based system is preferred over the IPU-initialised sentence-based system. The difference in performance of both systems is statistically significant ($p < 0.05$). The IPU-initialised sentence-based system still suffers from word repetitions in the synthesised output. 

\begin{table}[h!]
\centering
\caption{PC test results comparing Telugu IPU-initialised sentence and IPU-based (Tacotron2) systems}
\label{tab:chap5-pc_exp2_tel}
\begin{tabular}{|r|r|r|r|}
\hline
\multicolumn{1}{|c|}{\textbf{System}} & \multicolumn{1}{|c|}{\textbf{IPU-initialised sentence}} & \multicolumn{1}{c|}{\textbf{IPU}} & \multicolumn{1}{c|}{\textbf{Equal}} \\ \hline
\textbf{Preference (in $\%$}) & 14     &   69         &      17                          \\ \hline
\end{tabular}
\end{table}

The observations from the IPU initialisation of sentence-based systems are summarised below:

\begin{itemize}
    \item With the IPU initialisation, better alignments are obtained for the sentence-based system. This is mainly a consequence of better alignments learnt at the IPU level.
    \item The IPU initialisation leads to faster convergence when training with the sentence-level data.
    \item But even with better alignments, the initialised sentence-based system has many repetition errors during synthesis, compared to a purely IPU-based system. This indicates that the IPU approach is more suitable for synthesising conversational-style text.
\end{itemize}

\subsection{With non-autoregressive FastSpeech2 network}

Most E2E networks now include explicit duration modelling to reduce errors in the synthesised audio. In this context, we investigate the relevance of the IPU-based approach for non-autoregressive FastSpeech2 models \cite{fastspeech2}. Sentence and IPU-based FastSpeech2 models are trained for Hindi and Tamil datasets. Phone-level alignments are obtained for the training data using the Montreal Forced Aligner (MFA) \cite{MFA_interspeech17}.

\subsubsection{Error analysis}

Error analysis is carried out on the synthesised utterances generated by the sentence and IPU-based FastSpeech2 systems for Hindi. The models are trained as described in Sections \ref{sec:chap5-baseline} and \ref{sec:chap5-proposed_IPU}. The same set used in Table \ref{tab:chap5-error_analysis_Tac2} is considered for analysis. As expected, the IPU-based model does not have any repetition or word skip errors in the generated output. It is observed that even with the sentence-based FastSpeech2 model, there are no repetition or word skip errors in the generated output. This is mainly due to explicit duration modelling in FastSpeech2.

\subsubsection{Comparison of training time}

Table \ref{tab:chap5-training_time_FS2} shows the average training time per epoch for the FastSpeech2 systems. Compared to Tacotron2 systems (Table \ref{tab:chap5-training_time}), the gap between the training time for sentence and IPU-based FastSpeech2 systems has reduced. Despite this, an average relative reduction in training time by  $10.24\%$ is obtained with the IPU approach.

\begin{table}[h!]
\centering
\caption{Comparison of training time across different FastSpeech2 systems}
\label{tab:chap5-training_time_FS2}
\begin{tabular}{|l|cr|}
\hline
\multicolumn{1}{|c|}{\multirow{2}{*}{Dataset}} & \multicolumn{2}{c|}{Training time per epoch (min)}                   \\ \cline{2-3} 
\multicolumn{1}{|c|}{}                         & \multicolumn{1}{c|}{Sentence-based} & \multicolumn{1}{c|}{IPU-based} \\ \hline
Hindi male                                     & \multicolumn{1}{r|}{3.41}           & 3.13                           \\ \hline
Tamil female                                   & \multicolumn{1}{r|}{3.75}           & 3.29                            \\ \hline
\end{tabular}
\end{table}

\subsubsection{Subjective evaluation}

A PC test is conducted to evaluate the comparative performance between the different FastSpeech2 systems. From the set considered for Hindi error analysis, 20 synthesised utterances are randomly chosen. For Tamil, translated text segments of a classroom lecture on the subject ``Python Programming'' from NPTEL \cite{nptel} are synthesised. From this set, 20 utterances are selected. Preceding and succeeding utterances are included for content completion. 

14 native Hindi and 10 Tamil speakers participated in the evaluation, with each listener evaluating 10 audio pairs. Results of the PC test for Hindi and Tamil are presented in Tables \ref{tab:chap5-pc_exp3_hin} and \ref{tab:chap5-pc_exp3_tam}, respectively. It is clearly seen that the IPU-based system is preferred over the conventional sentence-based system. The difference in performance between both systems is statistically significant ($p < 0.05$). The IPU-based Tamil system has a higher preference compared to its Hindi counterpart. This is mainly due to longer Tamil utterances, for which the IPU-based system has provided prosodically pleasing audio compared to the slightly more monotonous audio generated by the corresponding sentence-based system.

\begin{table}[h!]
\centering
\caption{PC test results comparing Hindi sentence and IPU-based (FastSpeech2) systems}
\label{tab:chap5-pc_exp3_hin}
\begin{tabular}{|r|r|r|r|}
\hline
\multicolumn{1}{|c|}{\textbf{System}} & \multicolumn{1}{|c|}{\textbf{Sentence}} & \multicolumn{1}{c|}{\textbf{IPU}} & \multicolumn{1}{c|}{\textbf{Equal}} \\ \hline
\textbf{Preference (in $\%$}) &  27.68    &  51.78           & 20.54                               \\ \hline
\end{tabular}
\end{table}

\begin{table}[h!]
\centering
\caption{PC test results comparing Tamil sentence and IPU-based (FastSpeech2) systems}
\label{tab:chap5-pc_exp3_tam}
\begin{tabular}{|r|r|r|r|}
\hline
\multicolumn{1}{|c|}{\textbf{System}} & \multicolumn{1}{|c|}{\textbf{Sentence}} & \multicolumn{1}{c|}{\textbf{IPU}} & \multicolumn{1}{c|}{\textbf{Equal}} \\ \hline
\textbf{Preference (in $\%$}) &   16.25   &      70.00       &     13.75                           \\ \hline
\end{tabular}
\end{table}

\subsubsection{Pitch analysis}

As outlined earlier, one of the goals of this work was to assess if the IPU-based synthesis was closer to conversational-style speech. Studies by \cite{bhagyashree_asru_2021} show that conversational speech is characterised by a wide variation in pitch and syllable rate compared to read speech. As a preliminary study, we compared the distributions of utterance-level mean and standard deviation of pitch values extracted from audio files generated by sentence and IPU-based systems. We observed no significant differences in these distributions across both categories of voices. In this context, a later work has studied this further and demonstrated the need for adapting TTS systems to highly curated conversational speech data to generate expressive speech \cite{ishika_2023}.

% \begin{landscape}
\begin{figure}[h!]
 \centering
 \includegraphics[width = \linewidth]{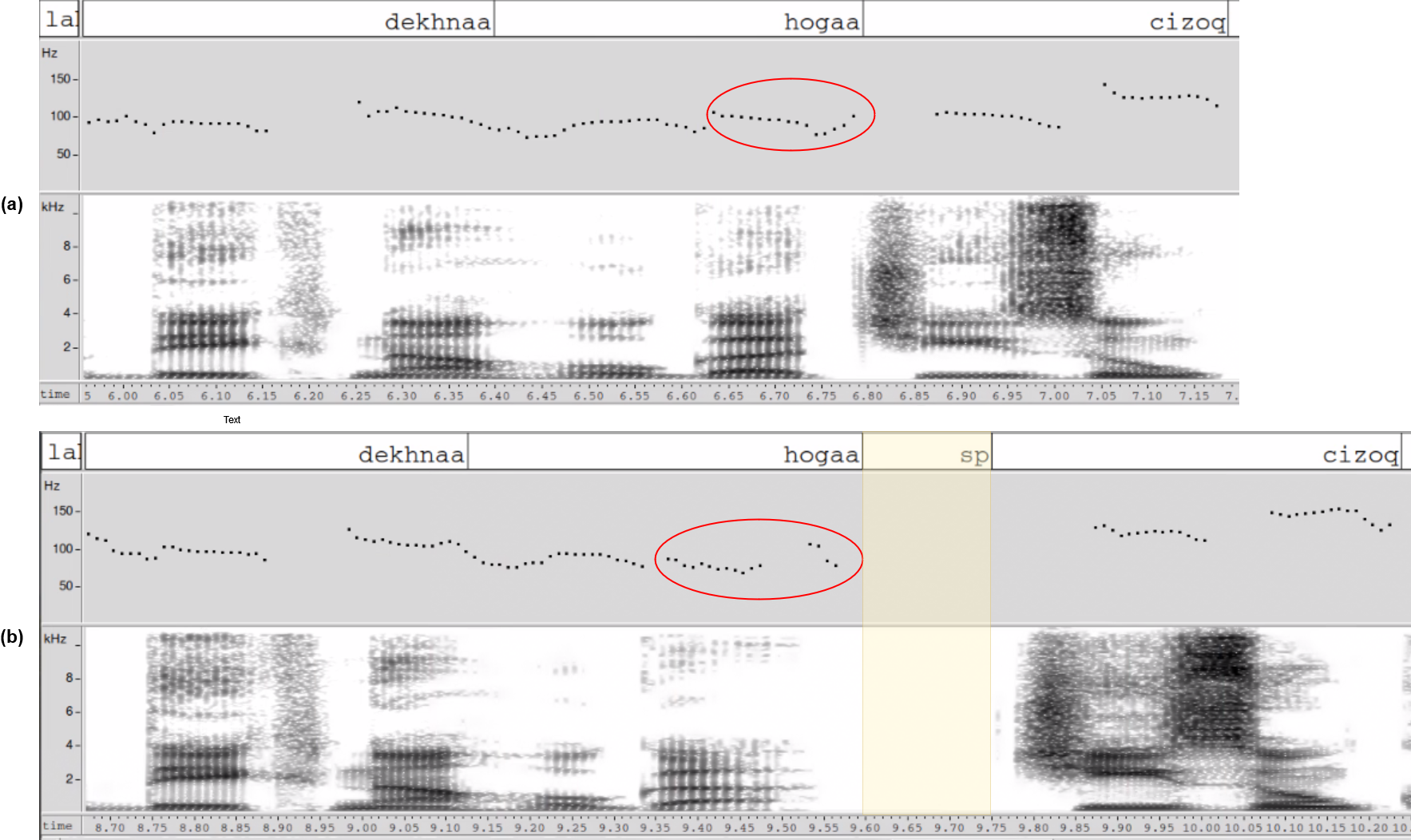}
   \caption{Pitch contours and spectrograms corresponding to audio files for a sample Hindi text synthesised by FastSpeech2 TTS systems: (a) sentence-based, (b) IPU-based.}
  \label{fig:chap5-pitch_exp3}
\end{figure}
% \end{landscape}

However, an interesting observation noticed while analysing individual synthesised utterances was that sometimes the sentence-based output did not provide the appropriate pitch contour. Consider the plots shown in Figure \ref{fig:chap5-pitch_exp3}. Each plot has three panels corresponding to the synthesised utterance-- word-level label, pitch contour and spectrogram. The utterance corresponds to a declarative-type text. In a declarative-type utterance, it is expected that the pitch falls towards the end of the phrase. The word ``hogaa'' indicates the end of the phrase. In Figure \ref{fig:chap5-pitch_exp3} (a), the region marked in red shows a rising pitch, thus indicating an incorrect pitch contour associated with the sentence-based synthesis. In contrast, the IPU-based synthesis has the correct pitch contour with a falling pitch at the end of the word ``hogaa'' (Figure \ref{fig:chap5-pitch_exp3} (b)). This indicates that there is better control over pitch resetting at the end of a phrase. Additionally, a short pause ``sp'' is generated by the IPU-based synthesis (highlighted in yellow) as a consequence of concatenating the IPU-level utterances. This, in turn, improves the comprehensibility of the utterance. Hence, considering the IPU as a unit for synthesis produces prosodically richer synthesis compared to sentence-based synthesis.

Our observations with the FastSpeech2 voices are summarised here:

\begin{itemize}
    \item With the FastSpeech2 sentence-based systems, the issue of word skips and repetitions has reduced owing to explicit duration modelling.

    \item The training time of sentence-based systems has reduced. However, IPU-based FastSpeech2 systems still require comparatively less resource time. 

    \item With the IPU-based synthesis, prosodically richer audio is generated compared to sentence-based synthesis. The generated output ensures pitch resetting between IPU segments and additionally includes a short silence region for better comprehensibility. In this context, identifying phrase boundaries is crucial.
    
\end{itemize}

\subsection{Auxiliary experiments, design choices and discussion}

In the experiments conducted, certain design choices were considered, and a few auxiliary experiments were carried out:

\begin{itemize}
    \item We experimented with different values of $T(sil)$= 100 msec, 200 msec, 300 msec, mean and mean $\pm$ standard deviation of intra-utterance silence duration. With a higher value of $T(sil)$, longer and fewer IPU segments were obtained. Preliminary investigations did not yield a conclusive pattern in performance with the different IPU models. 

    \item We also explored the hybrid segmentation technique \cite{BABY202010} for phone alignments in the context of IPU-based FastSpeech2 voices. Preliminary investigations found similar observations to those seen using MFA alignments. Further studies can be carried out in low-resource scenarios.
    
    \item In preliminary experiments with other Indic TTS datasets, the final synthesised utterance would sometimes be characterised by either inadequate pause or a lengthy pause duration at the points of IPU concatenation. For inadequate pause, we include a small silence region when joining individual synthesised IPU segments. We noticed that if the speaker in the training data has paused considerably, there is no need to include an extra silence region as the model learns this. Further investigation into how the speaking style of a speaker affects IPU synthesis is required.
    
    \item We included a comma as an IPU marker in the sentence-based Tacotron2 training. However, the model performance of this system was poor compared to that of the IPU-based voice, with issues of word skips and repetitions still present in the generated audio.
   
    \item IPU-based vocoders were trained. Based on informal evaluations, their performance was found to be on par with sentence-based vocoders. Hence, all experiments were conducted using sentence-based vocoders.

    \item In a few instances of IPU-based synthesis, the generated output was characterised by perceivable variability in energy, pitch and/or speed across the concatenated IPU segments. To some extent, this variability in prosody was reduced by constraining an IPU segment to have a minimum number of words. However, this may not always be feasible. A prospective solution may be to split the IPU training data into begin, middle and end IPU segments and train a TTS system separately for each category, as explored in \cite{Jeena_PhraseJournal_2019}, given more training data. Another solution would be to generate an utterance by super-imposing a specified prosody, especially in FastSpeech2.

    \item Experiments can also be explored by categorising the IPUs based on length or intra-utterance pauses. To this extent, a recent work \cite{duration_aware_ICASSP23} has improved the rhythm of the generated audio by employing pause phonemes based on categorised pause durations: brief ($<$ 300 ms), medium (300–700 ms), and long ($>$ 700 ms).
    
    \end{itemize}

\section{Summary}
\label{sec:summary}

In this work, we have explored an inter-pausal unit based approach for E2E speech synthesis. We have seen that the IPU-based autoregressive Tacotron2 training results in reduced word skip and repetition errors in the generated audio, compared to that in conventional sentence-based Tacotron2 systems. With non-autoregressive FastSpeech2 systems, the IPU-based approach results in lesser training time and prosodically richer synthesis. Since the approach explored in this work is architecture-independent, it can also be explored in the context of direct text-to-waveform architectures \cite{VITS_2021} and other emerging architectures.

\section*{Acknowledgment}

This work has been carried out as part of the following projects funded by different agencies: (1) ``Text to Speech Generation with chosen accent and noise profile for Aerospace and Industrial domains'' by the Department of Science and Technology (DST), Government of India (GoI), (2) ``Natural Language Translation Mission'' by the Ministry of Electronics and Information Technology (MeitY), GoI, (3) ``Speech to Speech Machine Translation'' by Office of the Principal Scientific Adviser (PSA), GoI, and (4) ``Speech Technologies in Indian Languages'' by MeitY, GoI.

\bibliographystyle{IEEEbib}
\bibliography{references}

\end{document}